\documentclass[aps,prd,twocolumn,showpacs,groupedaddress]{revtex4}  
\usepackage{graphicx}  
\usepackage{dcolumn}   
\usepackage{bm}        
\usepackage{amssymb}   
\usepackage{amsmath}
\usepackage{multirow}

\begin{document}
\newcommand{\dzero}{D0 }
\newcommand{\deteta}{\ensuremath{\eta_{D}}}
\newcommand{\effacc}{\ensuremath{\epsilon \times A}}
\newcommand{\Z}{\ensuremath{Z/\gamma^{*}}}


\hspace{5.2in} \leftline{Fermilab-Pub-07-040-E}

\title{Measurement of the shape of the boson rapidity distribution for $p\bar{p}\rightarrow Z/\gamma^{*}\rightarrow e^+e^- +X $ events produced at $\sqrt{s}$ of 1.96 TeV }
%
\author{                                                                      
V.M.~Abazov,$^{35}$                                                           
B.~Abbott,$^{75}$                                                             
M.~Abolins,$^{65}$                                                            
B.S.~Acharya,$^{28}$                                                          
M.~Adams,$^{51}$                                                              
T.~Adams,$^{49}$                                                              
E.~Aguilo,$^{5}$                                                              
S.H.~Ahn,$^{30}$                                                              
M.~Ahsan,$^{59}$                                                              
G.D.~Alexeev,$^{35}$
% ...
\\                                                                            
\vskip 0.30cm                                                                 
\centerline{(D\O\ Collaboration)}                                             
\vskip 0.30cm                                                                 
}                                                                             
\affiliation{                                                                 
\centerline{$^{1}$Universidad de Buenos Aires, Buenos Aires, Argentina}       
\centerline{$^{2}$LAFEX, Centro Brasileiro de Pesquisas F{\'\i}sicas,         
                  Rio de Janeiro, Brazil}                                     
\centerline{$^{3}$Universidade do Estado do Rio de Janeiro,                   
                  Rio de Janeiro, Brazil}                                     
\centerline{$^{4}$Instituto de F\'{\i}sica Te\'orica, Universidade            
                  Estadual Paulista, S\~ao Paulo, Brazil}                     
\centerline{$^{5}$University of Alberta, Edmonton, Alberta, Canada,           
                  Simon Fraser University, Burnaby, British Columbia, Canada,}
\centerline{York University, Toronto, Ontario, Canada, and                    
                  McGill University, Montreal, Quebec, Canada}                
\centerline{$^{6}$University of Science and Technology of China, Hefei,       
                  People's Republic of China}                                 
\centerline{$^{7}$Universidad de los Andes, Bogot\'{a}, Colombia}             
\centerline{$^{8}$Center for Particle Physics, Charles University,            
                  Prague, Czech Republic}                                     
\centerline{$^{9}$Czech Technical University, Prague, Czech Republic}         
\centerline{$^{10}$Center for Particle Physics, Institute of Physics,         
                   Academy of Sciences of the Czech Republic,                 
                   Prague, Czech Republic}                                    
\centerline{$^{11}$Universidad San Francisco de Quito, Quito, Ecuador}        
\centerline{$^{12}$Laboratoire de Physique Corpusculaire, IN2P3-CNRS,         
                   Universit\'e Blaise Pascal, Clermont-Ferrand, France}      
\centerline{$^{13}$Laboratoire de Physique Subatomique et de Cosmologie,      
                   IN2P3-CNRS, Universite de Grenoble 1, Grenoble, France}    
\centerline{$^{14}$CPPM, IN2P3-CNRS, Universit\'e de la M\'editerran\'ee,     
                   Marseille, France}                                         
\centerline{$^{15}$Laboratoire de l'Acc\'el\'erateur Lin\'eaire,              
                   IN2P3-CNRS et Universit\'e Paris-Sud, Orsay, France}       
\centerline{$^{16}$LPNHE, IN2P3-CNRS, Universit\'es Paris VI and VII,         
                   Paris, France}                                             
\centerline{$^{17}$DAPNIA/Service de Physique des Particules, CEA, Saclay,    
                   France}                                                    
\centerline{$^{18}$IPHC, IN2P3-CNRS, Universit\'e Louis Pasteur, Strasbourg,  
                   France, and Universit\'e de Haute Alsace,                  
                   Mulhouse, France}                                          
\centerline{$^{19}$IPNL, Universit\'e Lyon 1, CNRS/IN2P3, Villeurbanne, France
                   and Universit\'e de Lyon, Lyon, France}                    
\centerline{$^{20}$III. Physikalisches Institut A, RWTH Aachen,               
                   Aachen, Germany}                                           
\centerline{$^{21}$Physikalisches Institut, Universit{\"a}t Bonn,             
                   Bonn, Germany}                                             
\centerline{$^{22}$Physikalisches Institut, Universit{\"a}t Freiburg,         
                   Freiburg, Germany}                                         
\centerline{$^{23}$Institut f{\"u}r Physik, Universit{\"a}t Mainz,            
                   Mainz, Germany}                                            
\centerline{$^{24}$Ludwig-Maximilians-Universit{\"a}t M{\"u}nchen,            
                   M{\"u}nchen, Germany}                                      
\centerline{$^{25}$Fachbereich Physik, University of Wuppertal,               
                   Wuppertal, Germany}                                        
\centerline{$^{26}$Panjab University, Chandigarh, India}                      
\centerline{$^{27}$Delhi University, Delhi, India}                            
\centerline{$^{28}$Tata Institute of Fundamental Research, Mumbai, India}     
\centerline{$^{29}$University College Dublin, Dublin, Ireland}                
\centerline{$^{30}$Korea Detector Laboratory, Korea University,               
                   Seoul, Korea}                                              
\centerline{$^{31}$SungKyunKwan University, Suwon, Korea}                     
\centerline{$^{32}$CINVESTAV, Mexico City, Mexico}                            
\centerline{$^{33}$FOM-Institute NIKHEF and University of                     
                   Amsterdam/NIKHEF, Amsterdam, The Netherlands}              
\centerline{$^{34}$Radboud University Nijmegen/NIKHEF, Nijmegen, The          
                  Netherlands}                                                
\centerline{$^{35}$Joint Institute for Nuclear Research, Dubna, Russia}       
\centerline{$^{36}$Institute for Theoretical and Experimental Physics,        
                   Moscow, Russia}                                            
\centerline{$^{37}$Moscow State University, Moscow, Russia}                   
\centerline{$^{38}$Institute for High Energy Physics, Protvino, Russia}       
\centerline{$^{39}$Petersburg Nuclear Physics Institute,                      
                   St. Petersburg, Russia}                                    
\centerline{$^{40}$Lund University, Lund, Sweden, Royal Institute of          
                   Technology and Stockholm University, Stockholm,            
                   Sweden, and}                                               
\centerline{Uppsala University, Uppsala, Sweden}                              
\centerline{$^{41}$Physik Institut der Universit{\"a}t Z{\"u}rich,            
                   Z{\"u}rich, Switzerland}                                   
\centerline{$^{42}$Lancaster University, Lancaster, United Kingdom}           
\centerline{$^{43}$Imperial College, London, United Kingdom}                  
\centerline{$^{44}$University of Manchester, Manchester, United Kingdom}      
\centerline{$^{45}$University of Arizona, Tucson, Arizona 85721, USA}         
\centerline{$^{46}$Lawrence Berkeley National Laboratory and University of    
                   California, Berkeley, California 94720, USA}               
\centerline{$^{47}$California State University, Fresno, California 93740, USA}
\centerline{$^{48}$University of California, Riverside, California 92521, USA}
\centerline{$^{49}$Florida State University, Tallahassee, Florida 32306, USA} 
\centerline{$^{50}$Fermi National Accelerator Laboratory,                     
            Batavia, Illinois 60510, USA}                                     
\centerline{$^{51}$University of Illinois at Chicago,                         
            Chicago, Illinois 60607, USA}                                     
\centerline{$^{52}$Northern Illinois University, DeKalb, Illinois 60115, USA} 
\centerline{$^{53}$Northwestern University, Evanston, Illinois 60208, USA}    
\centerline{$^{54}$Indiana University, Bloomington, Indiana 47405, USA}       
\centerline{$^{55}$University of Notre Dame, Notre Dame, Indiana 46556, USA}  
\centerline{$^{56}$Purdue University Calumet, Hammond, Indiana 46323, USA}    
\centerline{$^{57}$Iowa State University, Ames, Iowa 50011, USA}              
\centerline{$^{58}$University of Kansas, Lawrence, Kansas 66045, USA}         
\centerline{$^{59}$Kansas State University, Manhattan, Kansas 66506, USA}     
\centerline{$^{60}$Louisiana Tech University, Ruston, Louisiana 71272, USA}   
\centerline{$^{61}$University of Maryland, College Park, Maryland 20742, USA} 
\centerline{$^{62}$Boston University, Boston, Massachusetts 02215, USA}       
\centerline{$^{63}$Northeastern University, Boston, Massachusetts 02115, USA} 
\centerline{$^{64}$University of Michigan, Ann Arbor, Michigan 48109, USA}    
\centerline{$^{65}$Michigan State University,                                 
            East Lansing, Michigan 48824, USA}                                
\centerline{$^{66}$University of Mississippi,                                 
            University, Mississippi 38677, USA}                               
\centerline{$^{67}$University of Nebraska, Lincoln, Nebraska 68588, USA}      
\centerline{$^{68}$Princeton University, Princeton, New Jersey 08544, USA}    
\centerline{$^{69}$State University of New York, Buffalo, New York 14260, USA}
\centerline{$^{70}$Columbia University, New York, New York 10027, USA}        
\centerline{$^{71}$University of Rochester, Rochester, New York 14627, USA}   
\centerline{$^{72}$State University of New York,                              
            Stony Brook, New York 11794, USA}                                 
\centerline{$^{73}$Brookhaven National Laboratory, Upton, New York 11973, USA}
\centerline{$^{74}$Langston University, Langston, Oklahoma 73050, USA}        
\centerline{$^{75}$University of Oklahoma, Norman, Oklahoma 73019, USA}       
\centerline{$^{76}$Oklahoma State University, Stillwater, Oklahoma 74078, USA}
\centerline{$^{77}$Brown University, Providence, Rhode Island 02912, USA}     
\centerline{$^{78}$University of Texas, Arlington, Texas 76019, USA}          
\centerline{$^{79}$Southern Methodist University, Dallas, Texas 75275, USA}   
\centerline{$^{80}$Rice University, Houston, Texas 77005, USA}                
\centerline{$^{81}$University of Virginia, Charlottesville,                   
            Virginia 22901, USA}                                              
\centerline{$^{82}$University of Washington, Seattle, Washington 98195, USA}  
}                                                                             
\date{February 14, 2007}

\begin{abstract}
We present a measurement of the shape of the boson rapidity 
distribution for $p\bar{p}\rightarrow Z/\gamma^{*}\rightarrow e^+e^- +X$ events 
at a center-of-mass energy of 1.96 TeV. 
The measurement is made for events with electron-positron
mass $71 < M_{ee} < 111$ GeV 
and uses 0.4 fb$^{-1}$ of data collected 
at the Fermilab Tevatron collider with the \dzero detector.  
This measurement significantly reduces
the uncertainties on the rapidity distribution
in the forward region compared with previous measurements.  
Predictions of NNLO QCD are found to agree well with the data over
the full rapidity range.  
\end{abstract}

\maketitle 


\section{\label{sec:intro} Introduction}

Kinematic distributions of \Z\ bosons produced 
in hadronic collisions provide a wealth of information on the fundamental 
interactions involved. 
At leading order, \Z\ bosons are produced through the annihilation of a quark 
and an anti-quark, with the partons in the proton and anti-proton
carrying momentum fractions $x_1$ and $x_2$, respectively. 
In turn, the rapidity of the boson, defined as 
$
y=\frac{1}{2} \ln\frac{E+p_L}{E-p_L}
$, 
where $E$ is the energy of the  boson and $p_{L}$ is the component of its 
momentum along the beam direction, is directly related to the momentum fractions by 
$$x_{1,2}=\frac{M_{\Z}}{\sqrt{s}}e^{\pm y}.$$
Here, $M_{\Z}$ is the mass of the 
boson, and $\sqrt{s}$ is the center of mass energy. 
These kinematic distributions can be precisely reconstructed when the 
boson decays leptonically since the leptons can be
accurately reconstructed, and the backgrounds to di-lepton 
final states are small. 
For low rapidity bosons, the 
leptons also have  small pseudorapidity, 
$$\eta = -\ln \left( \tan \left( \theta/2 \right) \right),$$ 
where $\theta$ is the polar angle and is measured relative to the 
proton beam.
%
High rapidity bosons are more likely to have initial 
states that have maximal $|x_1-x_2|$ for the incident partons. 

Although calculations are 
available at next-to-next-to-leading-order in QCD (NNLO) for $\text {d}\sigma/\text {d}y$ for 
$p\bar{p} \rightarrow \Z \rightarrow \ell \bar{\ell} +X $ 
\cite{Anastasiou:2003ds}, 
few measurements of the differential cross section exist \cite{Affolder:2000rx}. 
The forward rapidity region ($|y|>1.5$) 
suffers from a smaller 
cross section and lower acceptance than the central rapidity
region ($|y|<1.5$),
and has not yet been well tested. 
The forward region probes quarks with low $x$ 
and high 4-momentum transfer squared $Q^{2}$ ($Q^2 \approx M_{Z}^2$) 
as well as quarks with very large $x$. 
Parton distribution functions (PDFs) in this regime are mainly 
determined by jet cross section data, which have very different
experimental and theoretical systematic uncertainties than the electron measurements  
presented here, and by inclusive lepton scattering data taken mostly 
at much lower $Q^2$, which must be evolved to high momentum 
transfer scales using the DGLAP equations \cite{dglap}.


We measure the normalized differential cross section 
\begin{equation*}
\frac{1}{\sigma}\left( \frac{\text {d}\sigma}{\text {d}y}\right)_{i} =
\frac{(\effacc)_{\text {avg}}}{N^{\text {obs}}_{\text {total}} -N^{\text {bg}}_{\text {total}}}\frac{N^{\text {obs}}_i-N^{\text {bg}}_i}{\Delta_i\left(\effacc\right)_i},
\end{equation*}
where the index $i$ indicates the boson rapidity bin. 
In the first term on the right hand side, $\epsilon_{\text {avg}}$ is the average 
efficiency and $A_{\text {avg}}$ is the average acceptance for 
kinematic and geometric cuts. 
 $N^{\text {obs}}_{\text {total}}$ is the total number of 
candidate bosons, and $N_{\text {total}}^{\text {bg}}$ is the total 
number of background events 
in the sample.   
In the second term, $\epsilon_{i}$, 
$A_{i}$, $N^{\text {obs}}_{i}$, and $N^{\text {bg}}_{i}$ are the same as before, but determined in each bin $i$. 
$\Delta_{i}$ is the bin width. 
Dividing by 
the total number of events
reduces many of the systematic uncertainties,
particularly those due to luminosity. 


The \dzero detector \cite{run2det} allows efficient detection of electrons \cite{electrons} 
at the large 
pseudorapidities needed to study high rapidity \Z\ bosons. 
It has a central tracking system, consisting of a 
silicon microstrip tracker (SMT) and a central fiber tracker (CFT), 
both located within a 2~T superconducting solenoidal 
magnet, with designs optimized for tracking and 
vertexing \cite{vertexing} at pseudorapidities $|\deteta|<3$ and $|\deteta|<2.5$, respectively.
The quantity \deteta\ is $\eta$ measured from the center of the detector.
A liquid-argon and uranium calorimeter allows reconstruction
of electrons, photons, jets, and missing transverse energy.
The calorimeter is
divided into a 
central section (CC) covering  $|\deteta| \lesssim 1.1$ , and two end calorimeters (EC) that extend coverage 
to $|\deteta|\approx 4.2$. Each calorimeter is housed in a separate 
cryostat~\cite{run1det}. 
An outer muon system, covering $|\deteta|<2$, 
consists of a layer of tracking detectors and scintillation trigger 
counters in front of 1.8~T iron toroids, followed by two similar layers 
after the toroids~\cite{run2muon}. 
The luminosity is measured using plastic scintillator 
arrays placed in front of the EC cryostats. The trigger and data 
acquisition systems are designed to accommodate the high luminosities 
of Run II. 

\section{\label{sec:selection} Event Selection}
This measurement utilizes a data set of 0.4 fb$^{-1}$ collected 
at the Fermilab Tevatron between 2002 and 2004. The data are 
from $p\bar{p}$ collisions at $\sqrt{s}=1.96$ TeV.  
We consider 
candidate $Z/\gamma^{*}$ events that decay into an 
electron-positron pair 
with a reconstructed invariant mass  $71 < M_{ee} < 111$ GeV. The 
range used is $\pm 20$ GeV about the mass of the $Z$ boson.

To optimize the acceptance for electrons at large $\eta$, two 
strategies are used. The first is to require only   
one of the electrons be matched to a 
reconstructed track. Requiring a track-matched electron helps to 
reduce background from jets misidentified as electrons, while removing the 
track requirement on the second electron 
extends the $\eta$ coverage beyond that of the tracking system.
The second strategy takes advantage of the length of the 
bunches containing the incident protons and
antiprotons, which has a design length of 37 cm. For our dataset, the  
$z$ coordinate \cite{coord_syst} 
of the primary 
interactions have a Gaussian distribution with an rms ranging from 
$29 \pm 2$ cm to $24 \pm 1$ cm. The rms varies with respect 
to run conditions and time. 
At large values of vertex $|z|$, some of the decay 
products will travel back through the detector 
towards smaller $|z|$ values. These particles pass 
through much of the active volume of the tracking system.  
Typically, these events have low background and have
the highest boson rapidities.

Events are considered only if a single electron trigger 
fired. The efficiency is $(99.0 \pm 0.3)\%$ per 
electron for particles with transverse momentum $p_{T} > 30$ GeV and $|\eta_D| < 2$. 
From these events, di-electron candidates 
are selected by requiring  
two isolated electromagnetic objects which have shower shapes 
consistent with those of electrons. 
A candidate electron is considered to be isolated when 
greater than $85\%$ of its energy is contained within a 
cone of 0.2 in $\eta - \phi$ space. Also, they must have deposited 
more than $90\%$ of  their energy in the 
electromagnetic portion of the calorimeter. 
One electron must have $p_T > 15$ GeV while the 
other electron has $p_T > 25$ GeV.    
Electrons are defined to be in the CC (EC) region of the calorimeter 
if they are within $|\deteta| < 0.9$ ($1.5 < |\deteta| < 3.2$). 
In the CC region, electrons are not used if they pass near 
EM calorimeter module boundaries.  
As mentioned previously, 
at least one of the electrons in each di-electron pair 
must be spatially matched to a reconstructed track. 
In addition, all CC-region electrons are required to have a 
track match, so that both legs of a CC-CC event have a track match. 
A total of 19,306 events pass these selection criteria. 

\section{\label{sec:effic} Efficiencies and Backgrounds}

Single electron efficiencies are measured 
from this data sample using a  
``tag-and-probe method.'' 
This method involves selecting a sample of $Z \rightarrow e^+e^-$
candidate events by applying tight selection criteria to one of the
electron candidates, the ``tag leg,'' and very loose selection
criteria to the other electron candidate, ``the probe leg.''
The reconstructed mass of the tag and probe pair are required to be 
close to that of a $Z$ boson. 
The tag leg has tighter cuts to reduce 
the amount of background and to increase the probability 
that the event is really a boson decay event, and 
not from jets that are misidentified as electrons. 
The probe leg 
has looser cuts and is used to test the selection cut in question. 
While the efficiencies are measured with data, Monte Carlo events 
are used to test for 
biases in the efficiency measurements. For this purpose,  
$Z/\gamma^{*}$ Monte Carlo events are generated with {\sc pythia} \cite{pythia}  and 
are processed with a 
full \dzero detector simulation  based on
the {\sc geant} software package \cite{geant}, which models the 
interactions of particles with matter. 
Efficiencies are measured for identification of particles
like photons and electrons that shower in the electromagnetic 
calorimeter (``EM particles''), shower shape cuts, 
trigger, and track-matching probability.  All efficiencies are 
studied as a function of the \deteta\  of the probe electron. In 
addition, some of the efficiencies are measured with respect 
to additional quantities such as $p_T$ of the probe, 
vertex $z$ position of the event, boson $y$, 
or run number. 
Single electron efficiencies are relatively flat in \deteta\ for the 
CC region, 
and the values are typically larger than 90\%.     
In the 
EC region,  the efficiencies are sensitive to changes in the calorimeter 
geometry, to the finite coverage of the tracking system, and 
to the shape of the distribution of event vertices.  
Due to this sensitivity, effects of variations in the width of the vertex $z$
distribution of the course of Run~II are taken into 
account with a width-dependent efficiency. 

\begin{figure*}
\begin{tabular}{cc}
\includegraphics[width=0.46\linewidth]{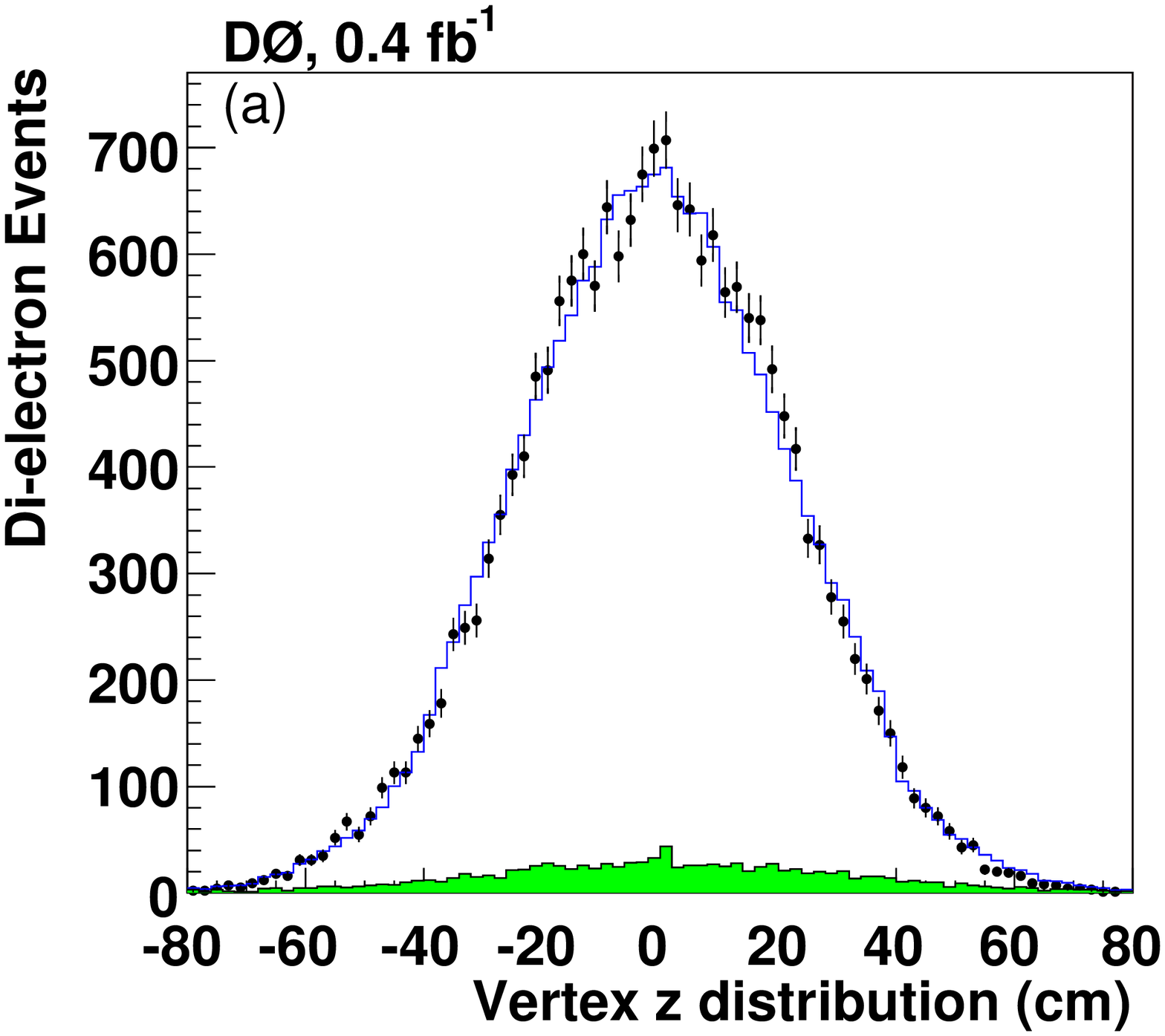} &
\includegraphics[width=0.46\linewidth]{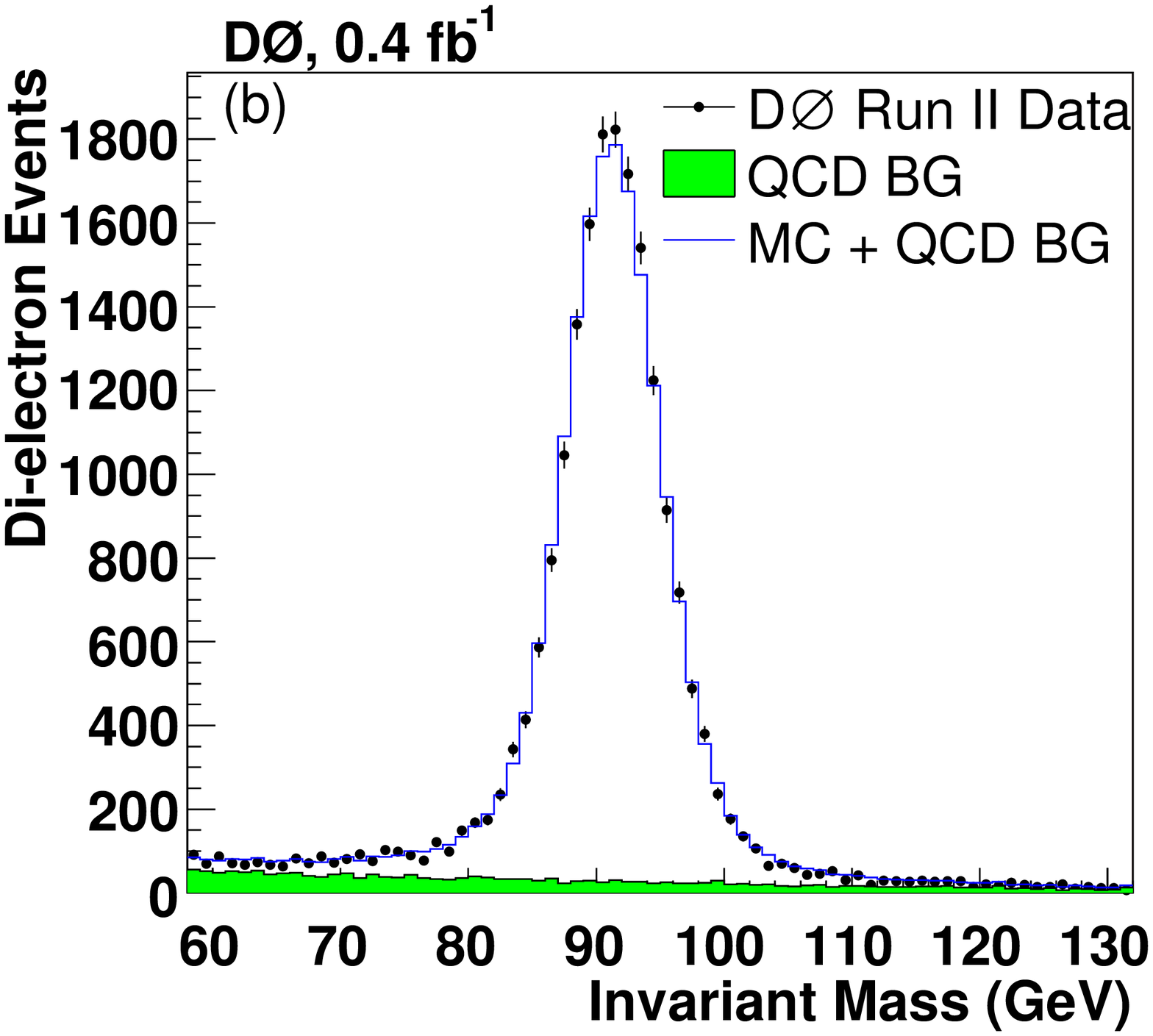} \\
\end{tabular}
\caption{\label{fig:data_mc_comparision_vtxz_and_mass}Comparisons of data 
and Monte Carlo plus background are presented for (a)
the vertex $z$ distribution and (b) the 
electron-positron invariant mass spectrum. The vertex $z$ plot 
shows data after all selection cuts. The data in the mass plot pass 
all selection criteria 
except for the mass cut. Uncertainties shown on the data points 
are statistical. }
\end{figure*}

For the acceptance 
determination, we use the {\sc resbos} Monte Carlo event 
generator \cite{Balazs:1997xd}  
with CTEQ6.1M input PDFs  \cite{cteq, Stump:2003yu}. {\sc resbos} computes the differential 
cross section including NLO QCD corrections and uses resummation for the 
low $p_T$  region. 
The simulated events then are processed with 
{\sc photos} \cite{Golonka:2005pn} to account for QED final 
state radiation (FSR).  
The events then are passed through a parameterized detector simulation which 
has been tuned to our data set. 
To properly apply efficiencies in the Monte Carlo,  
events are weighted based on the relative integrated luminosity 
per run. 
Figure~\ref{fig:data_mc_comparision_vtxz_and_mass} compares data
to the Monte Carlo simulation; the simulated signal plus
background reproduces the data well.
The \effacc\ per rapidity bin is summarized in Table~\ref{tab:results_tab_abs_y}.

The main source of background arises from events with jets where 
one or more of the jets are misidentified as an electron. 
The size of the background is less than 0.8\% for events where both 
electrons are detected in the central calorimeter (CC-CC) and less than 
6\% for the remainder of the data set. The background for CC-CC events is significantly 
smaller due to the track-match requirement on both electrons. 
The jet background is subtracted by fitting the 
di-lepton mass distribution with the sum of background and signal shapes.  
The signal shapes are taken from the same tuned Monte Carlo as used for 
the acceptance. 
Two different methods to determine the background shape are
used.
For $|y| < 2$, the background is determined separately for each 
rapidity bin. The total background is measured in the low statistics regions of $y>2$ and 
$y< -2$ separately. The small numbers of events in these areas do not 
permit the background fits to be performed on a bin-by-bin basis.   
In each of the high rapidity regions the background 
fraction is assumed to be constant for subsets of CC-CC, CC-EC, 
and EC-EC events.  
The background per bin then is determined  
using the number of candidate events per bin collected in each subset and the 
background fraction for that subset.

Additional background contributions could come from events 
that produce two 
EM objects in the final state. We consider diboson events 
containing a $W$ plus a $W$, $Z$, or $\gamma$;  
$t\bar{t}$;   and 
$Z/\gamma^{*}\rightarrow \tau \tau$ 
events where each $\tau$ decays to an electron.  The combined contribution 
from these additional sources is negligible compared to the 
background from jets. 
The total number of background events per bin is presented in Table~\ref{tab:results_tab_abs_y}. 

\section{\label{sec:systematics} Systematic Uncertainties}
A number of contributions to the systematic uncertainty 
are considered. These include contributions that arise from 
the determination of the \effacc\ correction and from the 
measurement of the background. 

For estimating the background systematic uncertainty, two different  background 
shapes are used in the fits. One is obtained from electron-positron events 
that fail the shower shape cuts; the other parameterizes 
the background shape as an exponential curve and incorporates it directly into the fit.
The exponential fits result in  about 13\% more background.   The average of the 
two methods is used as the background central value and the difference is split and assigned 
as a systematic uncertainty. An additional contribution to the 
background systematic uncertainty is 
derived by varying the constraints on the signal amplitude 
used in the background fits and redetermining 
the background.

The uncertainty on the differential cross section from the uncertainties on background 
ranges from $1.5 - 2.0\%$ for $|y|>0.8$. 
In this region the data mainly come from the CC-EC and EC-EC 
configurations which, because only one track-match is required,  
tends to allow more background. 
For $|y| < 0.8$, which is dominated by CC-CC data, the uncertainty 
due to the background 
is less than $1\%$. 

Several contributions due to the \effacc\ measurement are taken into account.   
These include the uncertainties on single 
electron efficiencies, the electron energy scale and energy resolution, 
the PDFs, and the model of the vertex $z$ distribution.

For single electron efficiencies, several aspects contribute to 
the systematic uncertainty. The first two contributions are 
derived from data while the third is obtained using Monte Carlo 
events. 
Since efficiencies are measured using  
data, the size of our $Z/\gamma^{*}$ sample inherently has a limited precision.  
This statistical uncertainty is included as part of the systematic uncertainty. 
The next component of the efficiencies' systematic uncertainty
comes from the background subtraction. To estimate 
this contribution, selection cuts are tightened on the tag electron to reduce 
the background at the expense of statistical precision. A comparison 
of the efficiencies with nominal and tighter cuts is used to estimate 
the systematic uncertainty from the background subtraction technique. 
Lastly, the tag-and-probe method used in the efficiency  measurement 
may produce a biased result if the efficiency for the 
probe electron passing the selection criteria is 
correlated with that of the tag electron. 
We estimate the size of this bias  with the {\sc pythia} 
Monte Carlo sample mentioned above, which includes a full detector simulation. 
Efficiencies measured using generator level information 
about the true particle identities 
are compared to the 
same efficiencies measured via tag-and-probe. 
The difference is used as a contribution to the systematic uncertainty. 

For the parameterized detector simulation,  
the energy response and resolutions are tuned using the width and peak position
of the electron-positron mass distribution
from the data sample.  Kinematic variables that are correlated with
the boson rapidity are not used to tune the detector simulation
parameters. 
Changing the tuning method leads to slight 
variations in the energy scale and resolution parameters. 
From these variations, we estimate the contribution due to uncertainties on the 
electron energy scale. The boson rapidity  measurement  
is not sensitive to  the energy resolution.   

CTEQ6.1M PDFs  are defined 
by twenty orthogonal parameters. 
Each parameter has an uncertainty which is shifted separately in the positive and negative 
direction to provide a set of forty PDFs
for  determination of the uncertainty. 
The acceptance is reevaluated with each PDF. 
Following the prescription presented in Ref.~\cite{Stump:2003yu},  
we compare each acceptance to that obtained with the nominal PDF set. 
The differences are combined into a PDF uncertainty, with a distinction made 
for sets that increase or decrease the acceptance.

As mentioned above, the shape of the vertex $z$ distribution varies with 
time. The width of the distribution can depend on a 
number of factors. The beam tuning has changed this width 
over the course of Run II. Also, the time elapsed since beam injection  
can affect the width.
Since the probability of an electron to have a track-match 
depends in part on the $z$ position of the primary vertex, knowledge 
of the vertex distribution can directly effect the acceptance correction. 
Samples of vertex $z$ distributions extracted from the data
in blocks corresponding to different instantaneous luminosities
are used to model the vertex distribution in the Monte Carlo
simulation.
Selection criteria 
that produce the widest and narrowest vertex distribution widths 
are used to estimate the systematic uncertainty. 

\begin{figure}
\includegraphics[scale=0.45]{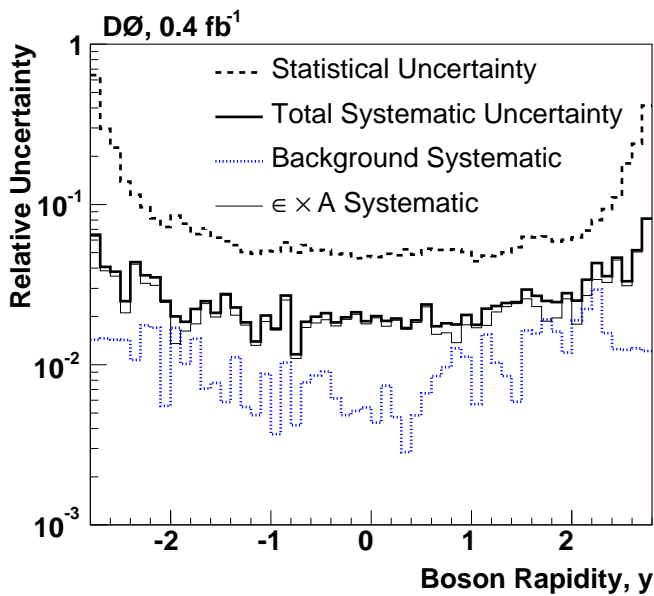}
\caption{\label{fig:select_rel_unc_v_y} Relative uncertainties plotted 
against the boson rapidity.} 
\end{figure}

The main contributions to the total systematic uncertainty  
depend on the boson rapidity. 
At small values of $|y|$, 
the main sources are the single electron efficiencies 
($\approx 2\%$) and the vertex $z$ distribution  ($<1\%$). 
For mid-range $|y|$, the largest contributions 
are due to the electron efficiencies and the background subtraction. The 
size of each is roughly $2\%$. In the high rapidity region, $|y|>2$, 
the main sources are from the electron efficiencies, the background, and 
the PDFs. The combined uncertainty in this region ranges from 
$ 3\%$ to $10\%$ and increases with $|y|$. 
The relative total systematic uncertainty 
along with the contributions to the uncertainties from the background and the \effacc\ 
are presented in Fig.~\ref{fig:select_rel_unc_v_y}. Contributions 
to the \effacc\ uncertainty also are presented for each rapidity bin in Table~\ref{tab:effacc_tab_abs_y}.

To cross check our result, we split the data into independent sets based on 
criteria that should not affect the result. 
These include  dividing the data based upon (a) time period for data collection, 
(b) different ranges of instantaneous luminosity, and 
(c) the calorimeter region in which the electrons are detected. 
Cross sections from independent subsets are 
compared to look for inconsistencies. Subsets in (a) are sensitive to hardware changes 
over the course of the data set and/or changes to the trigger menu used in collecting the data. 
Subsets in (b) have different vertex $z$ distributions and will not agree if the vertex 
distribution is modelled poorly. Subsets in group (c) compare data from three separate calorimeters. 
All of the cross checks give results that are consistent within uncertainties.

\begin{figure}
\includegraphics[scale=0.45]{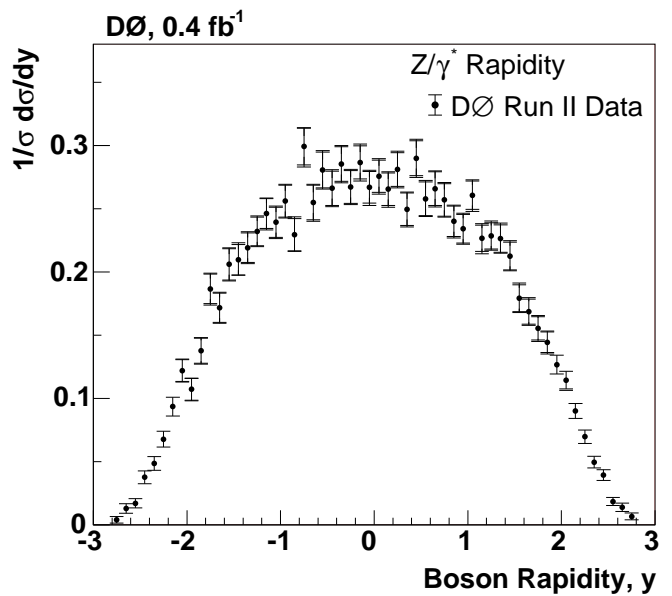}
\caption{\label{fig:1os_dsdy_v_y} D\O\ Run II measurement 
of $\frac{1}{\sigma} \text {d}\sigma/\text {d}y$ vs $y$. The inner (outer) error bars 
show the statistical (total) uncertainty.}
\end{figure}

\begin{figure}
\includegraphics[scale=0.45]{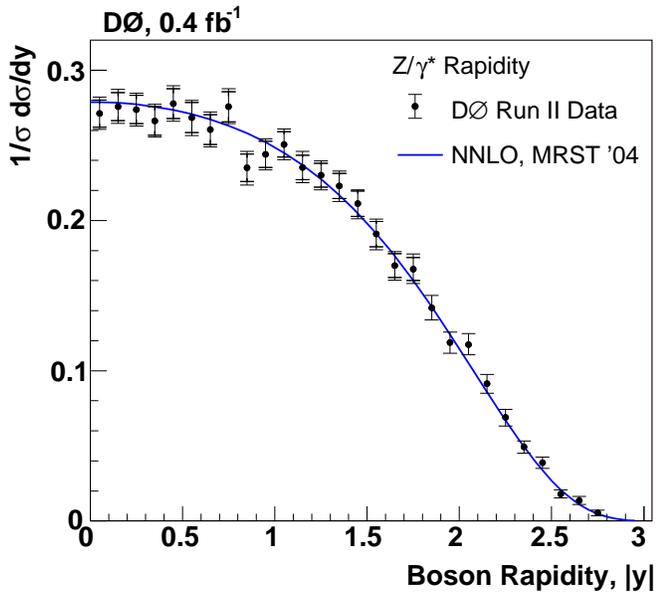}
\caption{\label{fig:1os_dsdy_v_abs_y} D\O\ Run II measurement 
of $\frac{1}{\sigma} \text {d}\sigma/\text {d}y$ vs $|y|$. The inner (outer) error bars 
show the statistical (total) uncertainty. The curve is a  
NNLO calculation from \cite{Anastasiou:2003ds} using MRST 2004 PDFs.}
\end{figure}

\section{\label{sec:results} Results}
A plot of  $\frac{1}{\sigma} \text {d}\sigma/\text {d}y$ is given 
in Fig.~\ref{fig:1os_dsdy_v_y} 
for $Z/\gamma^{*}$ events within a mass range of $71 <M_{ee} < 111$ GeV. 
The inner (outer) error bars show the statistical (total) uncertainty.
In Fig.~\ref{fig:1os_dsdy_v_abs_y} the result is shown vs $|y|$. 
For bin centering we follow the 
prescription given in Ref.~\cite{Lafferty:1994cj}. The center of the bin is located 
at the average value of the expected distribution over the bin. 
For this purpose we use 
the NNLO calculation generated from 
code made available from Ref.~\cite{Anastasiou:2003ds}. 
This is a small effect and for the two decimal places of precision used here, the procedure 
gives the same result as using the bin center. 
Due to the finite resolution of the \dzero detector, some fraction of 
the events in a given rapidity bin originates from a neighboring bin. 
For this analysis, 
about 5\% of the events migrate to each adjacent bin. 
Even though the effect is small, a migration correction is included in the 
\effacc\ determination. 
The uncertainties in $\frac{1}{\sigma}\text {d}\sigma/\text {d}y$ are dominantly
statistical for all measured rapidity bins.

\begin{figure}
\includegraphics[scale=0.45]{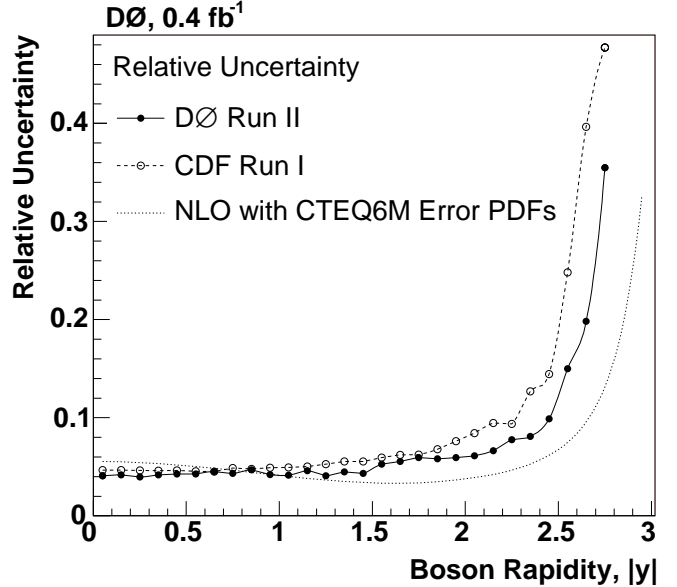}
\caption{\label{fig:rel_unc_d0_cdf_v_abs_y} Relative uncertainties from 
this measurement and from the CDF Run I result. Also shown is the 
PDF uncertainty on the differential cross section using the CTEQ6M 
uncertainty PDF sets. The values for the CTEQ6M curve are generated with 
code from Ref. \cite{Anastasiou:2003ds}. }
\end{figure}

The values for the fraction of the cross section in each rapidity bin
also are listed in Table~\ref{tab:results_tab_abs_y}. 
No information on the bin-to-bin correlations is included in the table.  
Since the systematic uncertainty is small compared to the 
statistical uncertainty, a correlation matrix is not included.
The curve  in Fig.~\ref{fig:1os_dsdy_v_abs_y} 
is a NNLO calculation from 
Ref.~\cite{Anastasiou:2003ds} generated with MRST 2004 
NNLO PDFs \cite{Martin:2004ir}. 
The calculation agrees well with our data,
with a 
$\chi^{2}/$d.o.f. of 20.0/27. 
Our result improves upon previous measurements over the full range
in $y$, especially in the forward region.
Figure~\ref{fig:rel_unc_d0_cdf_v_abs_y} shows the relative 
uncertainties from the most recent boson rapidity measurement
\cite{Affolder:2000rx}
and from this analysis plotted
vs $|y|$. For comparison purposes, Fig.~\ref{fig:rel_unc_d0_cdf_v_abs_y} 
also includes the relative uncertainty due to PDFs for a 
NLO calculation of the differential cross section. This curve 
uses CTEQ6M uncertainty PDFs and code from Ref.~\cite{Anastasiou:2003ds}. 
The method is the same as that used for the determination of 
the PDF uncertainty on the \effacc\ which was discussed earlier.

In summary, we have presented a measurement of $\frac{1}{\sigma}\text {d}\sigma/\text {d}y$ for 
$Z/\gamma^{*}$ measured with electron-positron
 events in the mass range $71<M_{ee}< 111$ GeV. 
The measurement is the most precise measurement to date.  It 
improves upon previous measurements and gives a 
significantly more precise
measurement of the boson rapidity distribution in
the high rapidity region which probes the small $x$, high 
$Q^2$ portion of the parton distribution
functions which is least constrained by other data. 
The fractional uncertainty in the highest rapidity bin
is reduced by 30\%. 
We find the result to be consistent with a recent NNLO calculation. 
The current measurement is performed with (10--20)\% of the expected Run II 
integrated luminosity.  
An improved result is foreseen with the inclusion of additional data, 
which will reduce the current still-dominant statistical uncertainty.

%
We are grateful to  Csaba Balazs for his help with {\sc resbos}.
We thank the staffs at Fermilab and collaborating institutions, 
and acknowledge support from the 
DOE and NSF (USA);
CEA and CNRS/IN2P3 (France);
FASI, Rosatom and RFBR (Russia);
CAPES, CNPq, FAPERJ, FAPESP and FUNDUNESP (Brazil);
DAE and DST (India);
Colciencias (Colombia);
CONACyT (Mexico);
KRF and KOSEF (Korea);
CONICET and UBACyT (Argentina);
FOM (The Netherlands);
PPARC (United Kingdom);
MSMT (Czech Republic);
CRC Program, CFI, NSERC and WestGrid Project (Canada);
BMBF and DFG (Germany);
SFI (Ireland);
The Swedish Research Council (Sweden);
Research Corporation;
Alexander von Humboldt Foundation;
and the Marie Curie Program.
%

\thispagestyle{plain}
\begin{table*}[b] 
\begin{center} 
\caption{ Summary of the measurement of $\frac{1}{\sigma} \frac{\text {d}\sigma}{\text {d}y}$ per rapidity bin for $Z/\gamma^{*}\rightarrow e^+e^-$ events with mass $71 < M_{ee} < 111$ GeV. 
\label{tab:results_tab_abs_y}}\begin{tabular}{c r | r r l | r | r l | r r} \hline \hline 
                  $|y|$  & &                   $\frac{1}{N} \times \frac{\text {d}N}{\text{d}y} $   &                   $\pm$ stat.  &                 $\pm  $ syst.    & \multicolumn{1}{c|}{Candidate Events } & \multicolumn{2}{c|}{Background Events}   & \multicolumn{2}{c} {                  $\epsilon \times A$ } \\  \hline        
 \,   0.05  & \,\,\,\, &   0.271  &  $\pm $ $  0.009 $   & $^{+   0.006}_{-   0.005}$& $   961 \, \, \, \hspace*{0.4in} $ & $ \hspace*{0.2in} \,  21.8 $ & $\pm $ $ 2.5 \, $ & $  0.176 $ & $^{+ 0.005}_{- 0.006} $ \\ 
 \,   0.15  & \,\,\,\, &   0.276  &  $\pm $ $  0.009 $   & $^{+   0.007}_{-   0.006}$& $   961 \, \, \, \hspace*{0.4in} $ & $ \hspace*{0.2in} \,  28.1 $ & $\pm $ $ 3.3 \, $ & $  0.172 $ & $^{+ 0.005}_{- 0.006} $ \\ 
 \,   0.25  & \,\,\,\, &   0.274  &  $\pm $ $  0.009 $   & $^{+   0.005}_{-   0.005}$& $   924 \, \, \, \hspace*{0.4in} $ & $ \hspace*{0.2in} \,  29.0 $ & $\pm $ $ 2.1 \, $ & $  0.166 $ & $^{+ 0.004}_{- 0.004} $ \\ 
 \,   0.35  & \,\,\,\, &   0.266  &  $\pm $ $  0.010 $   & $^{+   0.006}_{-   0.005}$& $   879 \, \, \, \hspace*{0.4in} $ & $ \hspace*{0.2in} \,  33.7 $ & $\pm $ $ 2.8 \, $ & $  0.161 $ & $^{+ 0.004}_{- 0.005} $ \\ 
 \,   0.45  & \,\,\,\, &   0.278  &  $\pm $ $  0.010 $   & $^{+   0.007}_{-   0.006}$& $   898 \, \, \, \hspace*{0.4in} $ & $ \hspace*{0.2in} \,  37.5 $ & $\pm $ $ 3.7 \, $ & $  0.158 $ & $^{+ 0.004}_{- 0.004} $ \\ 
 \,   0.55  & \,\,\,\, &   0.269  &  $\pm $ $  0.010 $   & $^{+   0.006}_{-   0.007}$& $   870 \, \, \, \hspace*{0.4in} $ & $ \hspace*{0.2in} \,  50.6 $ & $\pm $ $ 3.5 \, $ & $  0.155 $ & $^{+ 0.005}_{- 0.003} $ \\ 
 \,   0.65  & \,\,\,\, &   0.260  &  $\pm $ $  0.010 $   & $^{+   0.006}_{-   0.006}$& $   882 \, \, \, \hspace*{0.4in} $ & $ \hspace*{0.2in} \,  71.3 $ & $\pm $ $ 3.7 \, $ & $  0.159 $ & $^{+ 0.004}_{- 0.004} $ \\ 
 \,   0.75  & \,\,\,\, &   0.276  &  $\pm $ $  0.010 $   & $^{+   0.007}_{-   0.005}$& $   967 \, \, \, \hspace*{0.4in} $ & $ \hspace*{0.2in} \,  74.4 $ & $\pm $ $ 4.2 \, $ & $  0.164 $ & $^{+ 0.003}_{- 0.005} $ \\ 
 \,   0.85  & \,\,\,\, &   0.235  &  $\pm $ $  0.009 $   & $^{+   0.006}_{-   0.007}$& $   895 \, \, \, \hspace*{0.4in} $ & $ \hspace*{0.2in} \,  88.9 $ & $\pm $ $ 5.6 \, $ & $  0.175 $ & $^{+ 0.005}_{- 0.004} $ \\ 
 \,   0.95  & \,\,\,\, &   0.244  &  $\pm $ $  0.009 $   & $^{+   0.005}_{-   0.006}$& $   988 \, \, \, \hspace*{0.4in} $ & $ \hspace*{0.2in} \,  79.0 $ & $\pm $ $ 5.4 \, $ & $  0.190 $ & $^{+ 0.005}_{- 0.004} $ \\ 
 \,   1.05  & \,\,\,\, &   0.251  &  $\pm $ $  0.008 $   & $^{+   0.006}_{-   0.006}$& $  1095 \, \, \, \hspace*{0.4in} $ & $ \hspace*{0.2in} \,  75.2 $ & $\pm $ $ 3.8 \, $ & $  0.207 $ & $^{+ 0.005}_{- 0.006} $ \\ 
 \,   1.15  & \,\,\,\, &   0.235  &  $\pm $ $  0.008 $   & $^{+   0.007}_{-   0.006}$& $  1106 \, \, \, \hspace*{0.4in} $ & $ \hspace*{0.2in} \,  98.0 $ & $\pm $ $ 8.2 \, $ & $  0.218 $ & $^{+ 0.005}_{- 0.007} $ \\ 
 \,   1.25  & \,\,\,\, &   0.230  &  $\pm $ $  0.008 $   & $^{+   0.005}_{-   0.006}$& $  1060 \, \, \, \hspace*{0.4in} $ & $ \hspace*{0.2in} \,  83.7 $ & $\pm $ $ 5.5 \, $ & $  0.216 $ & $^{+ 0.006}_{- 0.005} $ \\ 
 \,   1.35  & \,\,\,\, &   0.223  &  $\pm $ $  0.008 $   & $^{+   0.005}_{-   0.006}$& $   965 \, \, \, \hspace*{0.4in} $ & $ \hspace*{0.2in} \,  94.6 $ & $\pm $ $ 4.3 \, $ & $  0.199 $ & $^{+ 0.006}_{- 0.005} $ \\ 
 \,   1.45  & \,\,\,\, &   0.211  &  $\pm $ $  0.008 $   & $^{+   0.004}_{-   0.005}$& $   793 \, \, \, \hspace*{0.4in} $ & $ \hspace*{0.2in} \,  60.0 $ & $\pm $ $ 2.4 \, $ & $  0.177 $ & $^{+ 0.006}_{- 0.003} $ \\ 
 \,   1.55  & \,\,\,\, &   0.191  &  $\pm $ $  0.008 $   & $^{+   0.005}_{-   0.006}$& $   694 \, \, \, \hspace*{0.4in} $ & $ \hspace*{0.2in} \,  69.5 $ & $\pm $ $ 5.4 \, $ & $  0.167 $ & $^{+ 0.005}_{- 0.004} $ \\ 
 \,   1.65  & \,\,\,\, &   0.170  &  $\pm $ $  0.008 $   & $^{+   0.005}_{-   0.005}$& $   644 \, \, \, \hspace*{0.4in} $ & $ \hspace*{0.2in} \,  72.4 $ & $\pm $ $ 5.1 \, $ & $  0.171 $ & $^{+ 0.006}_{- 0.005} $ \\ 
 \,   1.75  & \,\,\,\, &   0.168  &  $\pm $ $  0.008 $   & $^{+   0.006}_{-   0.006}$& $   689 \, \, \, \hspace*{0.4in} $ & $ \hspace*{0.2in} \,  79.5 $ & $\pm $ $ 6.3 \, $ & $  0.185 $ & $^{+ 0.005}_{- 0.006} $ \\ 
 \,   1.85  & \,\,\,\, &   0.142  &  $\pm $ $  0.007 $   & $^{+   0.005}_{-   0.004}$& $   614 \, \, \, \hspace*{0.4in} $ & $ \hspace*{0.2in} \,  57.3 $ & $\pm $ $ 5.4 \, $ & $  0.200 $ & $^{+ 0.005}_{- 0.007} $ \\ 
 \,   1.95  & \,\,\,\, &   0.119  &  $\pm $ $  0.006 $   & $^{+   0.004}_{-   0.004}$& $   559 \, \, \, \hspace*{0.4in} $ & $ \hspace*{0.2in} \,  55.5 $ & $\pm $ $ 3.9 \, $ & $  0.216 $ & $^{+ 0.006}_{- 0.006} $ \\ 
 \,   2.05  & \,\,\,\, &   0.117  &  $\pm $ $  0.006 $   & $^{+   0.005}_{-   0.004}$& $   551 \, \, \, \hspace*{0.4in} $ & $ \hspace*{0.2in} \,  37.3 $ & $\pm $ $ 5.7 \, $ & $  0.223 $ & $^{+ 0.007}_{- 0.009} $ \\ 
 \,   2.15  & \,\,\,\, &   0.091  &  $\pm $ $  0.005 $   & $^{+   0.004}_{-   0.004}$& $   459 \, \, \, \hspace*{0.4in} $ & $ \hspace*{0.2in} \,  36.4 $ & $\pm $ $ 5.6 \, $ & $  0.235 $ & $^{+ 0.010}_{- 0.007} $ \\ 
 \,   2.25  & \,\,\,\, &   0.069  &  $\pm $ $  0.004 $   & $^{+   0.003}_{-   0.004}$& $   352 \, \, \, \hspace*{0.4in} $ & $ \hspace*{0.2in} \,  33.2 $ & $\pm $ $ 5.7 \, $ & $  0.236 $ & $^{+ 0.011}_{- 0.008} $ \\ 
 \,   2.35  & \,\,\,\, &   0.049  &  $\pm $ $  0.004 $   & $^{+   0.002}_{-   0.002}$& $   232 \, \, \, \hspace*{0.4in} $ & $ \hspace*{0.2in} \,  15.4 $ & $\pm $ $ 2.1 \, $ & $  0.224 $ & $^{+ 0.012}_{- 0.008} $ \\ 
 \,   2.45  & \,\,\,\, &   0.039  &  $\pm $ $  0.003 $   & $^{+   0.002}_{-   0.002}$& $   162 \, \, \, \hspace*{0.4in} $ & $ \hspace*{0.2in} \,  10.4 $ & $\pm $ $ 1.1 \, $ & $  0.199 $ & $^{+ 0.010}_{- 0.012} $ \\ 
 \,   2.55  & \,\,\,\, &   0.018  &  $\pm $ $  0.003 $   & $^{+   0.001}_{-   0.001}$& $    61 \, \, \, \hspace*{0.4in} $ & $ \hspace*{0.2in} \,   3.9 $ & $\pm $ $ 0.4 \, $ & $  0.162 $ & $^{+ 0.008}_{- 0.011} $ \\ 
 \,   2.65  & \,\,\,\, &   0.014  &  $\pm $ $  0.003 $   & $^{+   0.001}_{-   0.001}$& $    35 \, \, \, \hspace*{0.4in} $ & $ \hspace*{0.2in} \,   2.2 $ & $\pm $ $ 0.2 \, $ & $  0.123 $ & $^{+ 0.008}_{- 0.012} $ \\ 
 \,   2.75  & \,\,\,\, &   0.005  &  $\pm $ $  0.002 $   & $^{+  0.0004}_{-  0.0004}$& $    10 \, \, \, \hspace*{0.4in} $ & $ \, \,   0.6 $              & $\pm $ $ 0.1 \, $ & $  0.085 $ & $^{+ 0.009}_{- 0.008} $ \\ 
\hline \hline \end{tabular} 
\end{center} 
\end{table*}

\pagestyle{empty} 
\begin{table*}[b] 
\begin{center} 
\caption{Contributions to the systematic uncertainty for $\epsilon \times A$ are shown in bins of $|y|$.   
Details of the contributions are described in the text.   
\label{tab:effacc_tab_abs_y}}\begin{tabular}{r | r | r | r | r | r | r | r } \hline \hline 
  $|y|$  & $ \epsilon \times A $   & $\delta $(total)   & \multicolumn{2}{c|}{$\delta $ ($e^-$  eff.) }   & $\delta $(PDF)   & $\delta $($E$ scale)   & $\delta $(vtx $z$)     \\         
    & \multicolumn{1}{c|}{ }  &   & \multicolumn{1}{c|}{stat. }   & \multicolumn{1}{c|}{method }  &   &   &    \\      \hline    
 0.05 &  0.176  & $^{+ 0.005 }_{- 0.006}$ & $\pm$ ${0.0004 }$ & $^{+ 0.004 }_{- 0.004}$ & $^{+ 0.002 }_{- 0.002}$ & $^{+0.0009 }_{-0.0002}$ & $^{+0.0016 }_{-0.0034}$ \\ 
 0.15 &  0.172  & $^{+ 0.005 }_{- 0.006}$ & $\pm$ ${0.0004 }$ & $^{+ 0.004 }_{- 0.004}$ & $^{+ 0.002 }_{- 0.002}$ & $^{+0.0010 }_{-0.0006}$ & $^{+0.0015 }_{-0.0033}$ \\ 
 0.25 &  0.166  & $^{+ 0.004 }_{- 0.004}$ & $\pm$ ${0.0004 }$ & $^{+ 0.004 }_{- 0.004}$ & $^{+ 0.002 }_{- 0.001}$ & $^{+0.0007 }_{-0.0011}$ & $^{+0.0012 }_{-0.0018}$ \\ 
 0.35 &  0.161  & $^{+ 0.004 }_{- 0.005}$ & $\pm$ ${0.0004 }$ & $^{+ 0.003 }_{- 0.004}$ & $^{+ 0.002 }_{- 0.003}$ & $^{+0.0002 }_{-0.0004}$ & $^{+0.0012 }_{-0.0017}$ \\ 
 0.45 &  0.158  & $^{+ 0.004 }_{- 0.004}$ & $\pm$ ${0.0004 }$ & $^{+ 0.003 }_{- 0.003}$ & $^{+ 0.002 }_{- 0.003}$ & $^{+0.0006 }_{-0.0002}$ & $^{+0.0017 }_{-0.0003}$ \\ 
 0.55 &  0.155  & $^{+ 0.005 }_{- 0.003}$ & $\pm$ ${0.0004 }$ & $^{+ 0.003 }_{- 0.003}$ & $^{+ 0.002 }_{- 0.002}$ & $^{+0.0015 }_{-0.0007}$ & $^{+0.0017 }_{-0.0002}$ \\ 
 0.65 &  0.159  & $^{+ 0.004 }_{- 0.004}$ & $\pm$ ${0.0004 }$ & $^{+ 0.003 }_{- 0.004}$ & $^{+ 0.002 }_{- 0.002}$ & $^{+0.0013 }_{-0.0002}$ & $^{+0.0007 }_{-0.0013}$ \\ 
 0.75 &  0.164  & $^{+ 0.003 }_{- 0.005}$ & $\pm$ ${0.0004 }$ & $^{+ 0.003 }_{- 0.003}$ & $^{+ 0.001 }_{- 0.003}$ & $^{+0.0011 }_{-0.0003}$ & $^{+0.0007 }_{-0.0013}$ \\ 
 0.85 &  0.175  & $^{+ 0.005 }_{- 0.004}$ & $\pm$ ${0.0004 }$ & $^{+ 0.003 }_{- 0.004}$ & $^{+ 0.004 }_{- 0.001}$ & $^{+0.0013 }_{-0.0004}$ & $^{+0.0008 }_{-0.0004}$ \\ 
 0.95 &  0.190  & $^{+ 0.005 }_{- 0.004}$ & $\pm$ ${0.0004 }$ & $^{+ 0.003 }_{- 0.004}$ & $^{+ 0.003 }_{- 0.001}$ & $^{+0.0002 }_{-0.0013}$ & $^{+0.0008 }_{-0.0005}$ \\ 
 1.05 &  0.207  & $^{+ 0.005 }_{- 0.006}$ & $\pm$ ${0.0004 }$ & $^{+ 0.004 }_{- 0.004}$ & $^{+ 0.002 }_{- 0.003}$ & $^{+0.0007 }_{-0.0005}$ & $^{+0.0022 }_{-0.0035}$ \\ 
 1.15 &  0.218  & $^{+ 0.005 }_{- 0.007}$ & $\pm$ ${0.0004 }$ & $^{+ 0.003 }_{- 0.005}$ & $^{+ 0.002 }_{- 0.003}$ & $^{+0.0014 }_{-0.0005}$ & $^{+0.0023 }_{-0.0036}$ \\ 
 1.25 &  0.216  & $^{+ 0.006 }_{- 0.005}$ & $\pm$ ${0.0004 }$ & $^{+ 0.004 }_{- 0.004}$ & $^{+ 0.003 }_{- 0.001}$ & $^{+0.0013 }_{-0.0004}$ & $^{+0.0024 }_{-0.0030}$ \\ 
 1.35 &  0.199  & $^{+ 0.006 }_{- 0.005}$ & $\pm$ ${0.0004 }$ & $^{+ 0.004 }_{- 0.003}$ & $^{+ 0.004 }_{- 0.002}$ & $^{+0.0008 }_{-0.0004}$ & $^{+0.0022 }_{-0.0028}$ \\ 
 1.45 &  0.177  & $^{+ 0.006 }_{- 0.003}$ & $\pm$ ${0.0004 }$ & $^{+ 0.005 }_{- 0.003}$ & $^{+ 0.003 }_{- 0.002}$ & $^{+0.0004 }_{-0.0009}$ & $^{+0.0017 }_{-0.0007}$ \\ 
 1.55 &  0.167  & $^{+ 0.005 }_{- 0.004}$ & $\pm$ ${0.0004 }$ & $^{+ 0.005 }_{- 0.003}$ & $^{+ 0.002 }_{- 0.002}$ & $^{+0.0012 }_{-0.0009}$ & $^{+0.0016 }_{-0.0006}$ \\ 
 1.65 &  0.171  & $^{+ 0.006 }_{- 0.005}$ & $\pm$ ${0.0005 }$ & $^{+ 0.004 }_{- 0.002}$ & $^{+ 0.003 }_{- 0.002}$ & $^{+0.0013 }_{-0.0007}$ & $^{+0.0008 }_{-0.0036}$ \\ 
 1.75 &  0.185  & $^{+ 0.005 }_{- 0.006}$ & $\pm$ ${0.0005 }$ & $^{+ 0.003 }_{- 0.004}$ & $^{+ 0.004 }_{- 0.002}$ & $^{+0.0009 }_{-0.0006}$ & $^{+0.0009 }_{-0.0039}$ \\ 
 1.85 &  0.200  & $^{+ 0.005 }_{- 0.007}$ & $\pm$ ${0.0006 }$ & $^{+ 0.002 }_{- 0.006}$ & $^{+ 0.003 }_{- 0.003}$ & $^{+0.0015 }_{-0.0008}$ & $^{+0.0030 }_{-0.0027}$ \\ 
 1.95 &  0.216  & $^{+ 0.006 }_{- 0.006}$ & $\pm$ ${0.0006 }$ & $^{+ 0.004 }_{- 0.004}$ & $^{+ 0.003 }_{- 0.004}$ & $^{+0.0010 }_{-0.0001}$ & $^{+0.0032 }_{-0.0029}$ \\ 
 2.05 &  0.223  & $^{+ 0.007 }_{- 0.009}$ & $\pm$ ${0.0008 }$ & $^{+ 0.004 }_{- 0.005}$ & $^{+ 0.004 }_{- 0.006}$ & $^{+0.0026 }_{-0.0002}$ & $^{+0.0024 }_{-0.0049}$ \\ 
 2.15 &  0.235  & $^{+ 0.010 }_{- 0.007}$ & $\pm$ ${0.0009 }$ & $^{+ 0.006 }_{- 0.004}$ & $^{+ 0.007 }_{- 0.003}$ & $^{+0.0012 }_{-0.0013}$ & $^{+0.0025 }_{-0.0051}$ \\ 
 2.25 &  0.236  & $^{+ 0.011 }_{- 0.008}$ & $\pm$ ${0.0010 }$ & $^{+ 0.008 }_{- 0.005}$ & $^{+ 0.005 }_{- 0.004}$ & $^{+0.0011 }_{-0.0014}$ & $^{+0.0055 }_{-0.0035}$ \\ 
 2.35 &  0.224  & $^{+ 0.012 }_{- 0.008}$ & $\pm$ ${0.0013 }$ & $^{+ 0.008 }_{- 0.006}$ & $^{+ 0.007 }_{- 0.003}$ & $^{+0.0010 }_{-0.0019}$ & $^{+0.0052 }_{-0.0033}$ \\ 
 2.45 &  0.199  & $^{+ 0.010 }_{- 0.012}$ & $\pm$ ${0.0017 }$ & $^{+ 0.004 }_{- 0.007}$ & $^{+ 0.008 }_{- 0.010}$ & $^{+0.0028 }_{-0.0000}$ & $^{+0.0028 }_{-0.0006}$ \\ 
 2.55 &  0.162  & $^{+ 0.008 }_{- 0.011}$ & $\pm$ ${0.0016 }$ & $^{+ 0.005 }_{- 0.007}$ & $^{+ 0.005 }_{- 0.007}$ & $^{+0.0017 }_{-0.0031}$ & $^{+0.0022 }_{-0.0005}$ \\ 
 2.65 &  0.123  & $^{+ 0.008 }_{- 0.012}$ & $\pm$ ${0.0015 }$ & $^{+ 0.004 }_{- 0.006}$ & $^{+ 0.005 }_{- 0.008}$ & $^{+0.0019 }_{-0.0044}$ & $^{+0.0038 }_{-0.0043}$ \\ 
 2.75 &  0.085  & $^{+ 0.009 }_{- 0.008}$ & $\pm$ ${0.0015 }$ & $^{+ 0.002 }_{- 0.006}$ & $^{+ 0.006 }_{- 0.005}$ & $^{+0.0048 }_{-0.0008}$ & $^{+0.0026 }_{-0.0029}$ \\ 
\hline \hline \end{tabular} 
\end{center} 
\end{table*}

\thispagestyle{plain}


\begin{thebibliography}{99}
\bibitem{vertexing} {Vertices are determined by using the highest 
$\sum \log p_T^{\text track}$ 
in a two-pass algorithm. First the vertices are calculated 
using a loose selection criteria. 
For the second pass, the found vertex position and errors 
are used as inputs for track selection and vertex fitting.}
\bibitem{coord_syst}{\dzero uses a right-handed coordinate system 
in which the $z$-axis points along the proton beam direction and the $y$-axis 
points upward. $\theta$ is 
the polar angle and $\phi$ is the azimuthal angle.}
\end{thebibliography}
\end{document}